\documentclass[pre,amsmath,amssymb,twocolumn,nofootinbib,floatfix]{revtex4-2}

\usepackage[colorlinks=true,linkcolor=blue,urlcolor=blue,citecolor=blue]{hyperref}
\usepackage{graphicx}
\usepackage{bm}
\usepackage{times}
\usepackage{color}

\newcommand{\sP}{s^\mathrm{p}}
\newcommand{\sD}{s^\mathrm{d}}
\newcommand{\kBT}{k_\mathrm{B} T}

\newcommand{\figref}[1]{\mbox{Fig.\hspace{0.25em}\ref{#1}}}
\newcommand{\Eqref}[1]{\mbox{Eq.\hspace{0.25em}\eqref{#1}}}
\newcommand{\Eqsref}[1]{\mbox{Eqs.\hspace{0.25em}\eqref{#1}}}

\newcommand{\rmd}{\mathrm{d}}

\newcommand{\vect}{\boldsymbol}

\graphicspath{{figures/},{}}

\renewcommand{\section}[1]{}
\renewcommand{\subsection}[1]{}
\renewcommand{\tableofcontents}{}

\begin{document}

\title{Physical interactions enable energy-efficient Turing patterns}

\author{Cathelijne ter Burg$^{1,*}$, David Zwicker$^{1,*}$}

\address{$^1$ Max Planck Institute for Dynamics and Self-Organization, Am Fa{\ss}berg 17, 37077 G\"{o}ttingen, Germany}

\begin{abstract}

Patterns are ubiquitous in nature, but how they form is often unclear.
Turing developed a seminal theory to explain patterns based on reactions that counteract the equalizing tendency of diffusion.
These reactions require continuous energy input since the system otherwise would proceed to equilibrium, but what systems are energy-efficient is currently unclear.
To address this question, we introduce a thermodynamically-consistent model of a Turing system.
We reveal that repulsive interactions between the stereotypical activator and inhibitor reduce energy requirements significantly.
Interestingly, efficient patterns occur for weak activity, albeit at reduced amplitude.
Our results suggest that physical interactions might be central in forming natural patterns.


\end{abstract} 

\maketitle

\tableofcontents

\section{Introduction}

Nature is full of interesting patterns, including nano-crystals~\cite{Fuseya2021}, complex tissues~\cite{RechoHallouHannezo}, and geophysical phenomena~\cite{Goehring2004}.
One prominent model to explain such patterns goes back to Alan Turing~\cite{Turing1952}, who showed that patterns emerge spontaneously in mixtures of diffusing and reacting chemicals.
These \emph{Turing patterns} form when a locally self-enhancing activator is counterbalanced by a faster-diffusing inhibitor that suppresses production globally~\cite{Vittadello2021,Kondo2010}.
Turing patterns require energy input, since diffusion would equalize all profiles in a closed system.
In biological systems, this energy input can come in the form of ATP molecules that are hydrolyzed to ADP to drive reactions.
Other patterns might rely on external energy input, e.g., light, vibrations, or wind.
However, the standard description of patterns, based on reaction-diffusion equations, does not explicitly include any energy input, which precludes a detailed analysis.
In particular, that approach cannot identify energy-efficient patterns.

In this letter, we introduce a thermodynamically consistent description of pattern formation inspired by Turing.
In this model, diffusive and reactive fluxes are driven by chemical potentials, which allows a natural inclusion of physical interactions between components.
Such physical interactions can facilitate patterns~\cite{Menou2023}, and they are inevitable in biological systems, in particular inside cells, where strong interactions lead to phase separation~\cite{Banani2017,Tsai2022,Liu2013,Choi2020,Dignon2020}.
We here show that interactions can lower the energy demands of creating patterns.

\section{Model}

To build our model, we consider a mixture of an activator~$A$ and an inhibitor~$I$ described by the respective volume fractions~$\phi_A(\vect r, t)$ and $\phi_I(\vect r, t)$.
The remaining fraction, $\phi_S = 1-\phi_A-\phi_I$, accounts for a solvent component $S$.
Their dynamics are governed by partial differential equations~\cite{Menou2023},
\begin{align}
	\partial_t \phi_i = \nabla \cdot  (D_i \phi_i \nabla \bar{\mu}_i) +  \sP_i + \sD_i
	\;, \qquad i = A, I
	\label{Eqn_non_ideal}
\end{align}
which comprises diffusive fluxes (first term on the right) and source terms~$\sP_i$ and $\sD_i$, which respectively describe the production and degradation of components~$i=A,I$. 
The diffusive fluxes are proportional to the diffusivities~$D_i$, where we focus on the case of large solvent mobility (section III C in the Supplementary Material~\cite{SI}).
These diffusive and reactive fluxes are driven by differences in the non-dimensional exchange chemical potentials, $\bar{\mu}_i = (\nu/\kBT) \delta F/\delta \phi_i$, where $\kBT$ is the relevant energy scale and $\nu$ denotes the molecular volume, which we take to be the same for all components.
The behavior of the molecules is governed by the free energy~$F$, and we consider a minimal model,
\begin{multline}
	\label{eqn:free_energy}
	F[\phi_A, \phi_I] = \frac{\kBT}{\nu} \int\Bigl[
		\phi_A \ln\phi_A + \phi_I \ln\phi_I + \phi_S \ln\phi_S
 \\
		+  \chi \phi_A \phi_I  + \frac{w^2}{2} \left( |\nabla \phi_A|^2 + |\nabla \phi_I|^2\right)
	\Bigr] \rmd V 
	\;, 
\end{multline}
which describes a non-ideal, incompressible, and isothermal ternary fluid of species $A$, $I$, and $S$.
The first line in \Eqref{eqn:free_energy} accounts for translational entropies of all components, which drives normal diffusion.
The term proportional to the Flory parameter~$\chi$ accounts for potential physical interactions between activator~$A$ and inhibitor~$I$, whereas we assume that the solvent $S$ interacts with  neither $A$ nor $I$.
Finally, the last term penalizes interfaces, which will then exhibit a typical width~$w$.
This free energy, known as Flory--Huggins free energy~\cite{Cahn1958,Safran2018,Rubinstein2003}, describes attraction between $A$ and $I$ for negative $\chi$, whereas the two species repel each other for positive $\chi$.
In particular, without reactions ($\sP_i=\sD_i=0$) and for equal average fraction~$\bar\phi$ of $A$ and $I$, minimizing $F$ implies that $A$ would phase separate from $I$ if $\chi>\bar\phi^{-1}$~\cite{Menou2023}.
In contrast, $\chi=0$ corresponds to the ideal diffusion commonly discussed for Turing patterns.


\begin{figure*}
    \includegraphics[width=\linewidth]{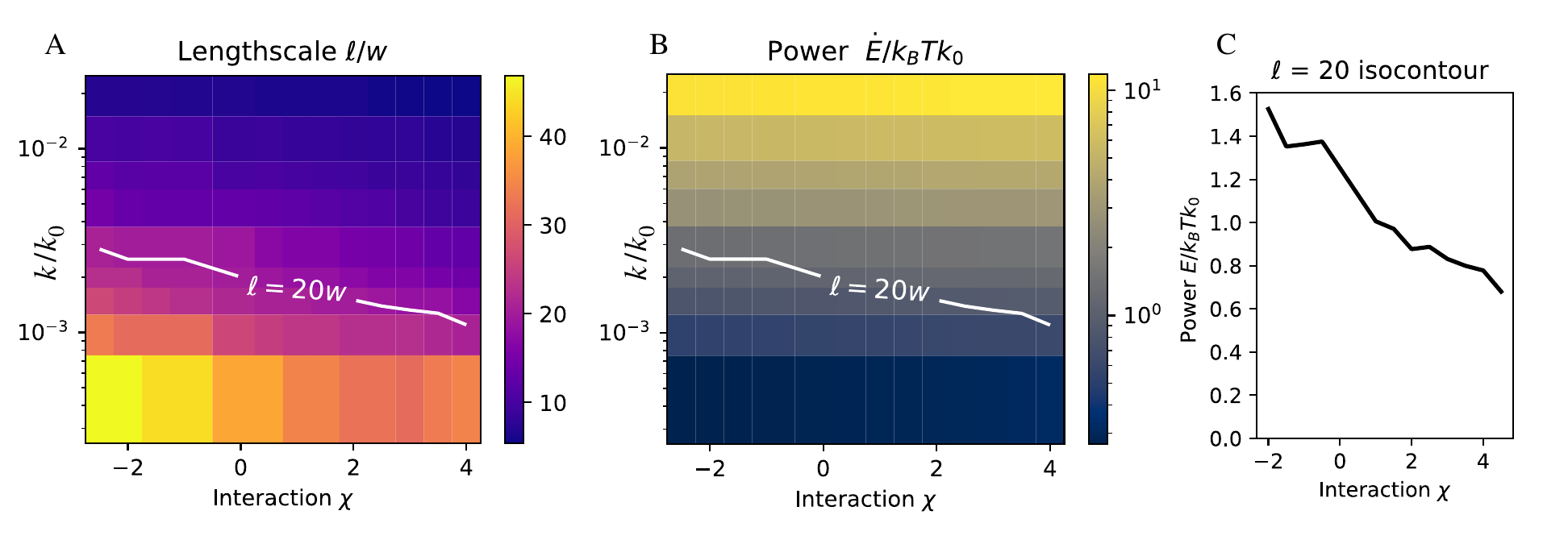}
    \caption{
    \textbf{Physical interactions reduce power requirements of Turing patterns.}
    (A) Pattern length scale $\ell$ determined from the number of  peaks in the stationary  profile $\phi_A(x)$ in a finite system as a function of interaction parameter $\chi$ and  reaction rate $k$. 
    (B) Power $\dot{E}$ given by Eq.~\eqref{Edot_eqn} as a function of $\chi$ and $k$.
	(C) $\dot{E}$ as a function of $\chi$ along the iso-contour for $\ell=20 w$ (white lines in A and B).
    (A--C) Model parameters are $\Delta \mu = 5$, $h = 5$, $D_I/D_A = 10$, $\phi_0 = 0.2$, and $k_0 = D_A/w^2$.  Simulations were performed on period grids of size $L = 2000\,w$.
	}
    \label{fig:Contoursdmu5} 
\end{figure*}


Turing patterns require non-linear reactions involving the activator~$A$ and inhibitor~$I$.
To analyze the energetics of our model, we choose thermodynamically consistent rate laws inspired by transition state theory~\cite{Atkin2010,Hanggi1990,Pagonabarraga1997,Bazant2013,Kirschbaum2021,Zwicker2022a}.
In brief, for a general reaction $X \rightleftharpoons Y$, the net reaction flux $s$ comprises two parts, $s=s_\rightarrow-s_\leftarrow$, stemming from the two directions in which the reaction can run.
These fluxes must obey a local detailed balance condition, $s_\rightarrow/s_\leftarrow = \exp[(\bar\mu_X - \bar\mu_Y)/\kBT]$, whereas absolute rates depend on kinetic details.
We use this freedom to express production and degradation of species $i=A,I$ as
\begin{subequations}
\label{eqn:reaction_fluxes}
\begin{align}
	\sP_i &= k   \frac{2 \phi_0}{1+(\phi_I / \phi_A)^h}\left[e^{\Delta \mu} - e^{\bar\mu_i}\right] ,   \label{Reaction_fluxes_prod}
\\
	\sD_i  &= k   \phi_i\left[e^{-\bar{\mu}_i} - e^{\Delta \mu}\right] , 
\label{Reaction_fluxes_degr}  
\end{align}
\end{subequations}
where the square bracket ensure the thermodynamic constraint of local detailed balance, and the pre-factor encodes the kinetics.
The thermodynamic part involves the exchange chemical potentials $\bar\mu_i$, which capture the energy required to convert solvent into species $i$.
To ensure that the system does not simply reduce to thermodynamic equilibrium, we also include 
a non-dimensional chemical potential difference~$\Delta\mu$, which describes a chemical driving force, e.g., provided by fuel that is kept available through a particle reservoir.
For $\Delta\mu=0$, the system obeys passive dynamics and relaxes toward equilibrium, which is characterized by $\bar\mu_i=0$, so that production and degradation cease separately ($\sP_i=\sD_i=0$) and patterns cannot be maintained.
To form patterns, we choose a non-trivial kinetic pre-factor, where $k$ is a rate constant determining the overall speed of the reactions, $\phi_0$ controls the average amount of $A$ and $I$ in the system, and the Hill coefficient $h$ quantifies the non-linearity in the production term~\cite{Weiss1997}, which biases production to regions where $\phi_A > \phi_I$.
The form of the kinetic pre-factor was chosen such that \Eqsref{eqn:reaction_fluxes} reduce to our previous model~\cite{Menou2023} in the limit of a strongly driven system ($\Delta\mu\rightarrow \infty$ with $k \rightarrow k e^{-\Delta\mu}$).
We thus expect regular patterns for sufficiently strong driving force~$\Delta\mu$, provided the other parameters (particularly $h$, $\phi_0$, $D_I/D_A$, and $\chi$) are suitably picked.

We analyze the dynamics given by \Eqsref{Eqn_non_ideal}--\eqref{eqn:reaction_fluxes} using numerical simulations and find patterns of various length scales as a function of $\chi$ and $k$ (\figref{fig:Contoursdmu5}A).
In particular, the observed length scale increases with decreasing~$k$, but it varies only weakly with $\chi$, consistent with previous results~\cite{Menou2023}.

\section{Results}

\subsection{Physical interactions reduce cost of maintaining patterns}

We leverage our thermodynamically consistent model to determine the energetic costs of maintaining a  periodic pattern. 
Energy is injected via the chemical driving force~$\Delta\mu$ described by \Eqref{eqn:reaction_fluxes} and dissipated via diffusive and reactive fluxes.
In stationary state, energy injection balances dissipation (see Supplementary Material, Section S4~\cite{SI}), allowing us to focus on the rate of injected energy,
\begin{equation}
	\dot{E} = \frac{1}{V}  \sum_{i = A, I} \int  (\sP_i - \sD_i)\Delta \mu \, \mathrm{d}V
	\;, \label{Edot_eqn}
\end{equation} 
which is the power required to maintain the pattern.
\figref{fig:Contoursdmu5}B shows that~$\dot E$ increases strongly with larger reaction rate~$k$, whereas the interaction~$\chi$ has a weaker effect. 
Combining \figref{fig:Contoursdmu5}A and \figref{fig:Contoursdmu5}B, we find that patterns with longer periods~$\ell$ are less  expensive, presumably because they require a weaker drive to counteract the equalizing tendency of diffusion.

Our model permits multiple parameter sets that lead to the same pattern length scale~$\ell$.
Are some of these parameters energetically favourable?
To analyze this in detail, we next determine the power~$\dot E$ at fixed pattern period~$\ell$.
\figref{fig:Contoursdmu5}C shows that $\dot E$ generally decreases with increasing $\chi$ along the isocontour of fixed $\ell$, which we also observe for other parameters (Fig. S1 in Supplementary Material \cite{SI}).
Taken together, this indicates that repulsion between the activator~$A$ and the inhibitor~$I$ can lower the required energy input to maintain a pattern of a given period~$\ell$. 


\subsection{Efficient patterns exhibit anti-correlated profiles}
\label{sec:CD-analytics}

\begin{figure}
	\includegraphics[width=\columnwidth]{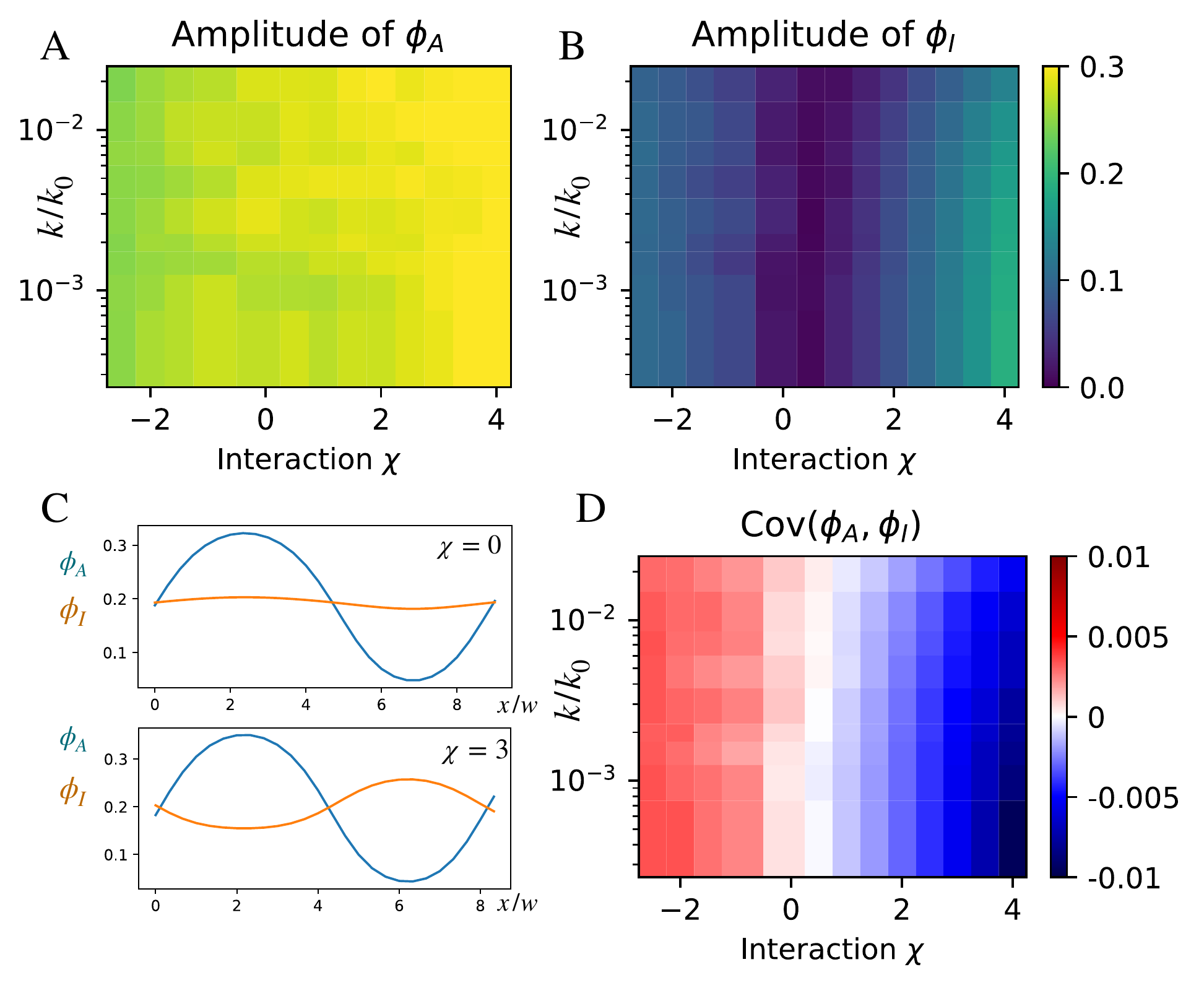}
	\caption{
	\textbf{Interactions induce anti-correlation between activator and inhibitor.}
	(A, B) Amplitude $\max(\phi_i) - \min(\phi_i)$ of activator profile $\phi_A$ (panel A) and inhibitor profile $\phi_I$ (panel B) for the data in \figref{fig:Contoursdmu5}. 
	(C) Activator profile $\phi_A$ (blue) and inhibitor profile $\phi_I$ (orange) as a function of space $x$ for two interaction parameters~$\chi$.
	(D) Covariance $\langle \phi_A \phi_I \rangle - \langle \phi_A  \rangle \langle  \phi_I \rangle $ for the data in \figref{fig:Contoursdmu5}.
	}
	\label{fig:Sim_Amp_Cov_profileDmu5}
\end{figure}

To understand why physical repulsion can lower the required power~$\dot E$, we analyze patterns in detail.
Physical attraction could help amplifying patterns, which would imply larger amplitudes, but the activator amplitude increases only slightly with the interaction strength~$\chi$ and is independent of the reaction rate $k$ (\figref{fig:Sim_Amp_Cov_profileDmu5}A).
In contrast to the activator, the amplitude of the inhibitor is vanishingly small, except for very large $\chi$ (\figref{fig:Sim_Amp_Cov_profileDmu5}B).
In particular, the inhibitor profile is almost flat without interactions ($\chi=0$, \figref{fig:Sim_Amp_Cov_profileDmu5}C), while the activator~$A$ shows pronounced variations.
This shape is consistent with the stereotypical mechanism of "local activation and global inhibition", which underlies Turing patterns~\cite{Vittadello2021,Kondo2010}.
However, this qualitative behavior deviates for larger interactions, where both species show significant variations (Fig. \ref{fig:Sim_Amp_Cov_profileDmu5}C).
In particular, the profiles are now anti-correlated, whereas they are slightly correlated for $\chi=0$.
Indeed, the covariance between $A$ and $I$ becomes more negative for larger $\chi$ (Fig. \ref{fig:Sim_Amp_Cov_profileDmu5}D), suggesting that anti-correlated patterns are associated with a lower cost.

\subsection{Anti-correlated profiles emerge from cross-diffusion}
We next examine how anti-correlated patterns can emerge and why they can lower the cost~$\dot E$ by analyzing the system for weak repulsive  interactions ($|\chi|< 5$), where phase separation is absent~\cite{Menou2023}. Assuming small variations of the profiles, we linearize the diffusion term in \Eqref{Eqn_non_ideal}, to obtain

\begin{align}
  \partial_t \phi_i = D_{ij} \nabla^2 \phi_j + \sP_i + \sD_i
  \;,
  \label{Eqn_non_ideal_crossdiff}
\end{align}
where the reaction terms are still given by \Eqsref{eqn:reaction_fluxes} and we introduced the constant diffusivity matrix
\begin{equation}
  D_{ij} = \begin{pmatrix}
   D_{A}(1+ \psi) & D_{A}(\psi + \chi \phi_*) \\ 
   D_{I}(\psi + \chi \phi_*) & D_{I}(1+ \psi)
  \end{pmatrix},
  \label{Dijcrossdiff}
\end{equation}
with $\psi = \phi_*/(1 - 2\phi_*)$. Here, $\phi_*$ denotes the concentration in the homogeneous steady state, which we obtain from the condition $\sP_i + \sD_i = 0$, using $\phi_A = \phi_I = \phi_*$ and the associated chemical potentials.
Numerical simulations of  \Eqref{Eqn_non_ideal_crossdiff} reproduce the essential features of the patterns of the full model  (Fig. S3 in Supplementary Material \cite{SI}), suggesting that the simplified model captures all essential features. 

The diffusivity matrix~$D_{ij}$ of the simplified model given by \Eqref{Dijcrossdiff} reveals that the interaction parameter~$\chi$ amplifies cross-diffusion since $\chi$ only appears in the off-diagonal terms.
In systems without interactions ($\chi = 0$) and $\psi \ll 1$, $D_{ij}$ is essentially diagonal, implying that the flux of each species is governed predominantly by its own concentration gradient.
In contrast, if $\chi \approx \phi_*^{-1}$, the off-diagonal terms are comparable to the diagonal terms, implying strong coupling of the diffusion of the two species, suggesting that this cross-diffusion could explain the reduced cost.
However, large $\chi$ also influence the chemical potentials, and thus the reaction fluxes given by \Eqsref{eqn:reaction_fluxes}, which might also explain the reduced cost. 
To test this alternative hypothesis, we next consider the limits $\Delta \mu \to \infty$ and $h \to \infty$, where the reactions fluxes reduce to piecewise constant functions, which only depend on $k$, $\phi_0$, and $\phi_i$, but are independent of $\chi$.
This case can be solved analytically (Section S3 in Supplementary Material~\cite{SI}), and the resulting profiles are  consistent with numerical simulations of  \Eqref{Eqn_non_ideal_crossdiff}  (Fig.~S4 in Supplementary Material \cite{SI}).
We thus conclude that the influence of $\chi$ onto the reaction rates is negligible, and instead cross-diffusion is the major factor allowing to form patterns at reduced cost.

\begin{figure}
	\centering
	\includegraphics[width=\columnwidth]{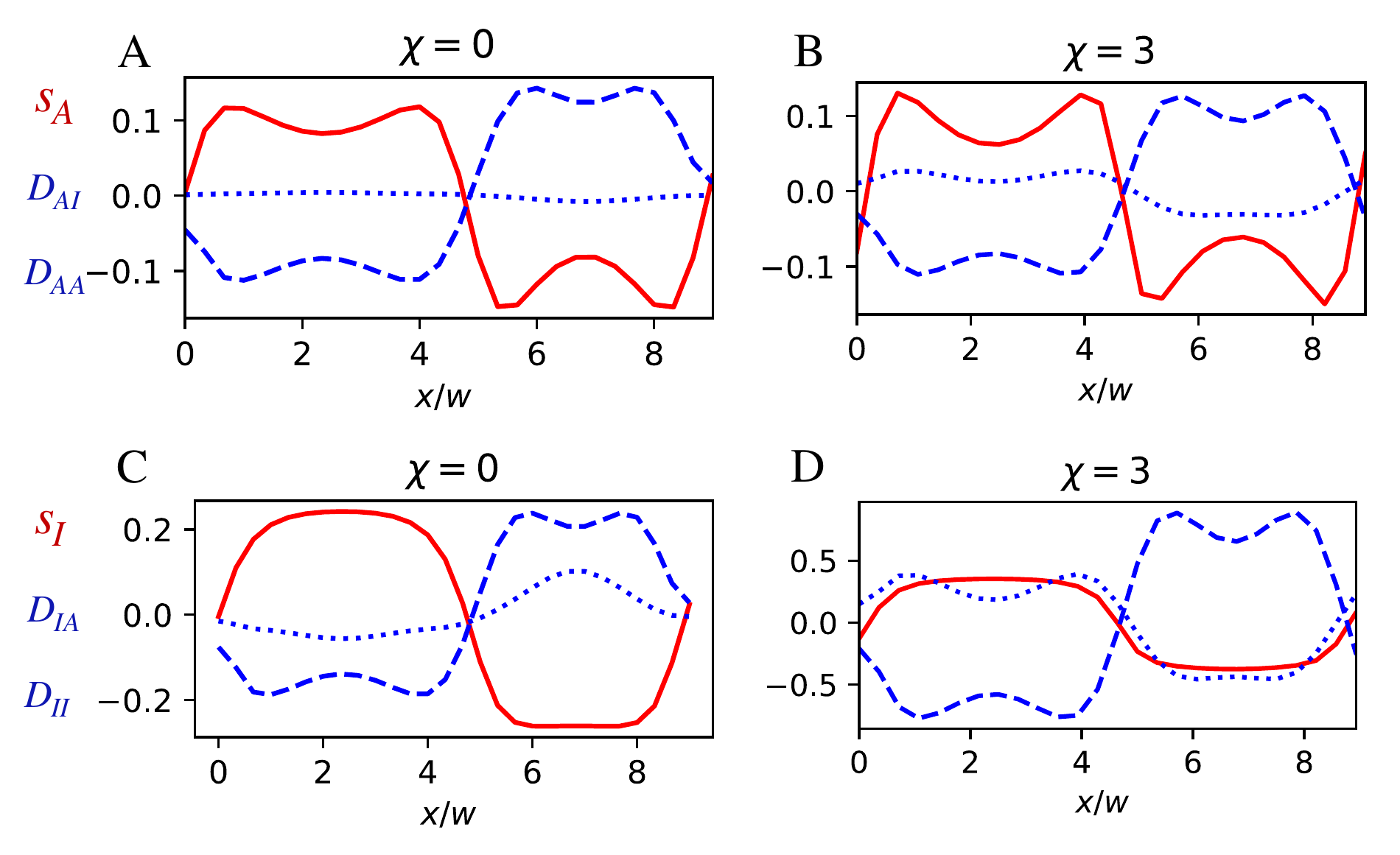}
	\caption{
	\textbf{Cross-diffusion aids pattern formation.}
	Spatial profiles of the three terms on the right hand side of \Eqref{Eqn_CrossDiffExplicit} for the activator (upper row) and inhibitor (lower row) as a function of space~$x$ for two interaction parameters~$\chi$ for the data in \figref{fig:Contoursdmu5}, $k= 0.01$. 
	The equalizing self-diffusion (blue dashed lines) is often opposed by cross-diffusion (blue dotted lines) and the reactions (red lines).
	}
	\label{fig:FluxDiff}
\end{figure}

To understand how cross-diffusion affects patterns, it is instructive to write \Eqref{Eqn_non_ideal_crossdiff} as
\begin{align}
	\partial_t \phi_i = D_{iA} \nabla^2 \phi_A +  D_{iI} \nabla^2 \phi_I + s_i
	\;,  \label{Eqn_CrossDiffExplicit}
\end{align}
for $i=A,I$ with the total reaction flux  $s_i = \sP_i + \sD_i$.
Analyzing the three terms on the right hand side individually confirms that cross-diffusion is essentially absent without interactions ($\chi=0$, \figref{fig:FluxDiff}A), implying that  self-diffusion must balance reactions in the stationary state.
In contrast, strong interactions ($\chi=3$, \figref{fig:FluxDiff}B) lead to significant cross-diffusion, which is correlated with the reaction flux.
Consequently, the equalizing self-diffusion is opposed by cross-diffusion and the reaction flux, allowing for a weaker reaction flux (and thus lower cost $\dot E$; see \Eqref{Edot_eqn} and section S4 in the Supplementary Material~\cite{SI})  to maintain similar profiles. 
Essentially, cross-diffusion, driven by the repulsion between activator and inhibitor, tends to separate the two components, which aids pattern formation.

\subsection{Varying $\Delta \mu$}
\label{s:varyDeltaMu}

\begin{figure}
	\includegraphics[width=\columnwidth]{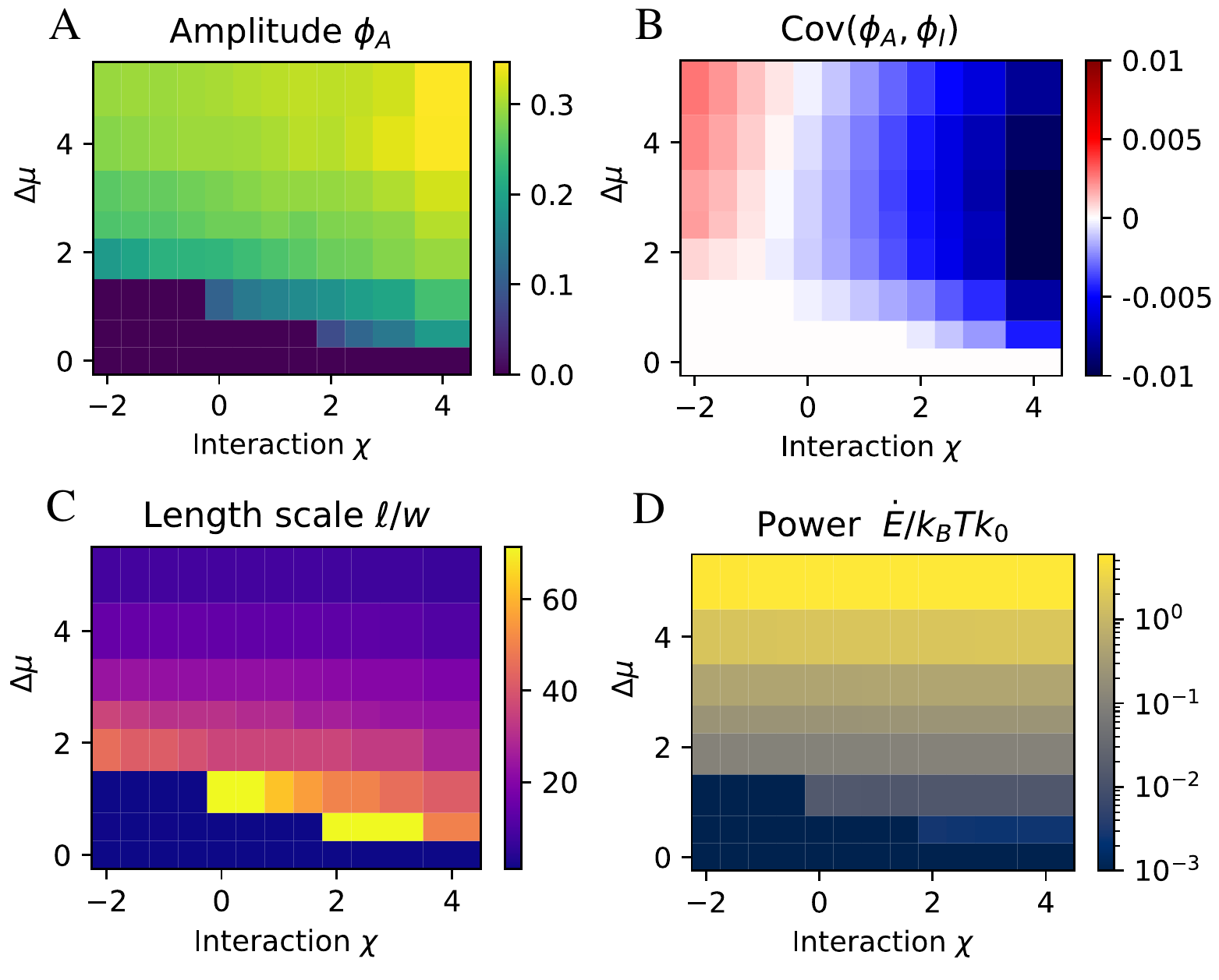}
	\caption{
	\textbf{Weak drive leads to efficient patterns with small amplitudes.} 
	Activator amplitude ($\max(\phi_A) - \min(\phi_A)$, panel A), covariance  between activator and inhibitor ($\langle \phi_A \phi_I \rangle - \langle \phi_A  \rangle \langle  \phi_I \rangle $, panel B), pattern length scale $\ell$ (panel C), and power $\dot E$ (\Eqref{Edot_eqn}, panel D) as functions of interaction $\chi$ and chemical driving force~$\Delta\mu$. Simulation parameters are $k = 0.01\, k_0$, $h = 8$, $D_I/D_A = 15$, $\phi_0 = 0.2$, and $k_0 = D_A/w^2$.
	Simulations were performed on periodic grids of size $L = 2000\,w$. 		}
	\label{fig_varydmu}
\end{figure}

So far, we have shown that interactions aid pattern formation for a particular driving force ($\Delta\mu=5$), and our thermodynamically consistent model implies vanishing patterns for vanishing drive ($\Delta\mu=0$).
To understand the detailed influence of $\Delta\mu$, we next consider fixed reactions rates~$k$ since this value has little influence on the amplitudes and covariance of profiles (\figref{fig:Sim_Amp_Cov_profileDmu5}).
\figref{fig_varydmu} shows that there is a minimal value of $\Delta\mu$ required for pattern formation.
This minimal value decreases with increasing interaction strength~$\chi$, consistent with our result that repulsive interactions aid pattern formation.
If $\Delta\mu$ is sufficiently large, it has little influence on the amplitude, covariance, and pattern length scale (\figref{fig:Sim_Amp_Cov_profileDmu5}A--C), similar to the influence of $k$.
However, larger $\Delta\mu$ strongly increase the power $\dot E$ (\figref{fig:Sim_Amp_Cov_profileDmu5}D), suggesting that there is an optimal value for $\Delta\mu$ that minimizes $\dot E$ for a given length scale~$\ell$.
Our data suggests that this optimal value is close to the threshold value for pattern formation.
Indeed, we find that smaller $\Delta\mu$ leads to larger average composition $\phi_*$ (Fig. S6 in Supplementary Material \cite{SI}), implying larger $\psi$, and thus larger cross-diffusion (\Eqref{Dijcrossdiff}).
This effect also explains the weak anti-correlation between $A$ and $I$ that we observe for small $\Delta\mu$ even at $\chi=0$ (\figref{fig:Sim_Amp_Cov_profileDmu5}B).
Taken together, the increased cross-diffusion  at small $\Delta\mu$ explains why pattern formation requires only small powers~$\dot E$ (\figref{fig:Sim_Amp_Cov_profileDmu5}D), but it comes at the cost of a reduced pattern amplitude (\figref{fig:Sim_Amp_Cov_profileDmu5}A) (also Section S2 in Supplementary Material~\cite{SI}).  

\section{Conclusion and discussion}


In summary, we have introduced a thermodynamically consistent model of a pattern forming mixture of an activator and an inhibitor. 
We found that repulsion between the components significantly reduced the energy required to maintain patterns, indicating that such physical interactions are crucial for pattern formation.
Essentially, the physical repulsion separates the inhibitor from the activator, which manifests in significant cross-diffusion.
Cross-diffusion has been identified as a key factor in many pattern forming systems~\cite{Jorne1974,Shigesada1979,Malchow1988,Malchow1990,Malchow1995,Vanag2009,Zemskov2011,Gaffney2022,Diez2023}, but the non-linearities that eventually lead to phase separation amplify the effect.
In contrast to standard Turing patterns, where molecules are produced in a restricted zone (local activation) and degraded everywhere (global inhibition), repulsion allows to concentrate the inhibitor (local inhibition), and thus anti-correlated patterns.

More generally, our analysis shows that the heterogeneous distribution of molecules, which is key for pattern formation, involves multiple processes.
Traditional reaction-diffusion models emphasize non-linear chemical reactions to create localized zones, but we showed that non-linear diffusive fluxes originating from physical interactions can play similar roles.
Since this diffusion does not require energy input, and dissipation mainly takes place via reactions in our case and other examples~\cite{Kumar2021}, it can lower the overall power required to maintain patterns.
Taken to the extreme, one can then use the physical interactions to produce droplets, and use chemical reactions to modify their behavior.
Such chemically active droplets indeed exhibit pattern formation~\cite{ZwickerPaulinTerBurg2025,Sastre2024,Julicher2024,Weber2019,Zwicker2022a,ZwickerOstwald2015,Soeding2020,Kirschbaum2021}, and it will be interesting to explore the rich physics of reacting non-ideal systems.

\bibliographystyle{apsrev4-1}
\bibliography{references}

\end{document}


\title{Supplementary Information for "Physical interactions enable energy-efficient Turing patterns"}

\author{Cathelijne ter Burg$^{1,*}$, David Zwicker$^{1,*}$}

\address{$^1$ Max Planck Institute for Dynamics and Self-Organization, Am Fa{\ss}berg 17, 37077 G\"{o}ttingen, Germany}

\maketitle

\onecolumngrid

\tableofcontents

\section{Details of numerical simulations}

We perform numerical simulations of Eq. (1) in the main text on an equidistantly discretized grid with periodic boundary conditions using second-order finite differences to approximate differential operators~\cite{Zwicker2020python}.
We evaluate $\nabla \mu_i$ on a staggered grid to ensure material conservation.
Numerical simulations of Eq. (5) in the main text also use an equidistantly discretized grid with second-order finite differences to approximate differential operators, but there is no need for a staggered grid.
In both cases, we used an explicit Euler scheme for the time evolution with adaptive timestepping. 

We typically use a spatial discretisation of two grid points per unit length $w$ and increased it to four when necessary.  
Finally, length scales reported in the main text are measured using the mean of the structure factor $S_q$, which gives a precise measurement for periodic patterns.
All simulations are performed in the limit of an infinitely fast solvent. 

\section{Additional results for $\Delta\mu=1$}
\label{s:WeakDrive}

In the main text, we show the behaviour of our model for $\Delta \mu = 5$.
To emphasize that the conclusions drawn in the main text are robust, we here present data for $\Delta \mu = 1$.
\figref{fig:Sim_Cov_profileDmu1} shows that in this case the amplitude increases strongly with larger interactions, and that patterns are anti-correlated for $\chi= 0$.
This suggests that while physical interactions still help to decrease the cost of pattern formation, they also promote stronger patterns and more strongly affect the details of the patterns, which is consistent with our conclusions about weak drive in the main text.

\begin{figure}
	\includegraphics[width=\columnwidth]{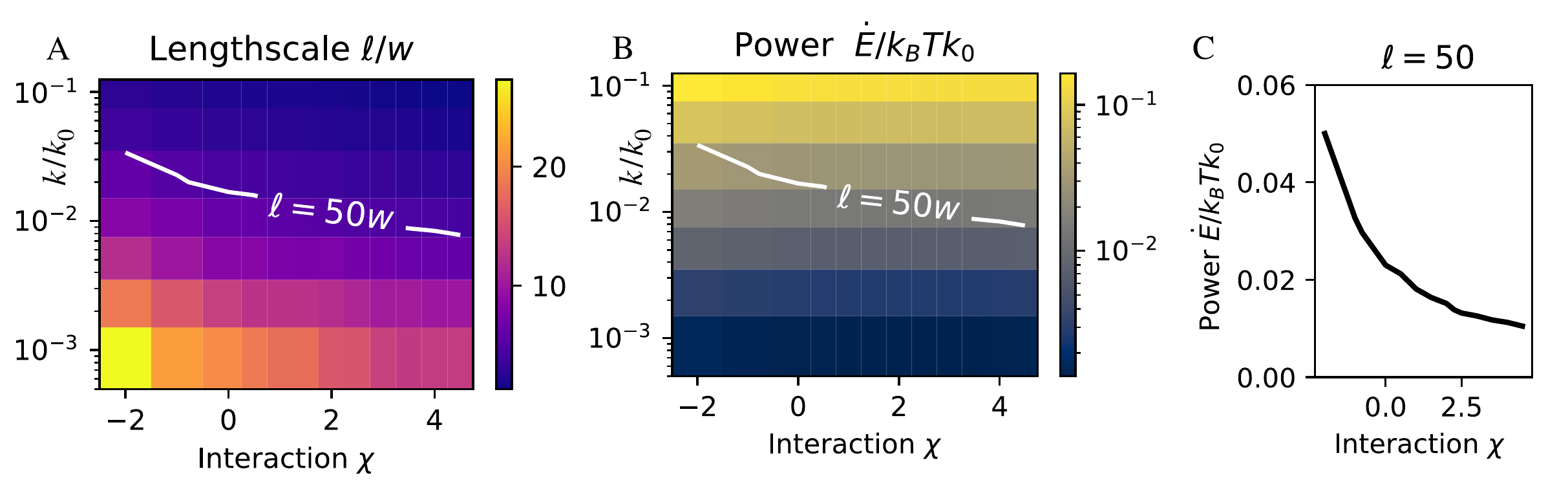}
	\caption{(A) Pattern length scale $l$ determined from the number of periodic peaks for $\Delta \mu = 1$. White contour in shows an iso-contour of fixed pattern length scale.
    (B) Power $\dot{E}$ measured from Eq. 4 for $\Delta \mu = 1$, with isocontour superimposed.
      (C) Power $\dot{E}$ along the iso-contours for $\Delta \mu = 1$ measured through Eq. 4. 
    (A--C)  Additional model parameters are  $h = 12$, $D_I/D_A = 20$, $k_0 = D_A/w^2, \phi_0 = 0.2$. Simulations where run on a periodic grid of size $L = 2000w$ to $L=   10000w$ depending on the prior estimated  pattern length scale and to ensure at least 10-20 periods to minimise discretisation errors.   }
	\label{fig:Sim_Amp_profileDmu1}
\end{figure}

\begin{figure}
	\includegraphics[width=\columnwidth]{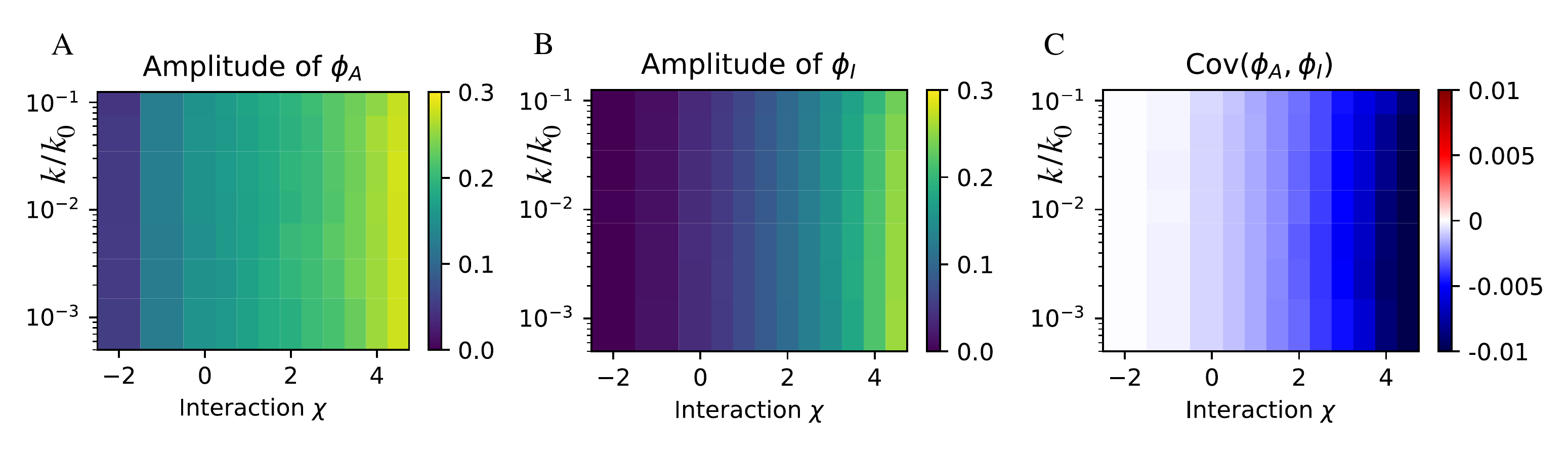}
	\caption{
		(A, B) Amplitude $\max(\phi_i) - \min(\phi_i)$ of activator and inhibitor as a function of interaction strength $\chi$ and reaction rate $k$.
		(C) Covariance $\langle \phi_A \phi_I \rangle - \langle \phi_A  \rangle \langle  \phi_I \rangle $ between activator and inhibitor  as a function of $\chi$ and $k$.
		(A--C) Data corresponding to Fig. \ref{fig:Sim_Amp_profileDmu1}. }
	\label{fig:Sim_Cov_profileDmu1}
\end{figure}

\newpage 

\section{Exactly solvable limit of $\Delta \mu \to \infty$, $h \to \infty$}
\label{App-crossDiff}

In the limit $\Delta \mu \to \infty$ and $h \to \infty$, it is possible to calculate the profiles analytically.
We first discuss ideal diffusion in section \ref{app_Analytical_idealDiff} and then cross-diffusion in  section \ref{app_Analytical_crossDiff}. 

\subsection{Ideal diffusion}
\label{app_Analytical_idealDiff}
 
In the limit $h \to  \infty$, the production reaction term becomes a Heaviside step function that switches between $2 k \phi_0 $ when $\phi_A > \phi_I$ and $0$ when $\phi_I > \phi_A$. 
Consequently, we can define a production and degradation region with respective domain sizes $L_\mathrm{p}$ and $L_\mathrm{d}$.
The dynamics in each region are described by
\begin{subequations}
\begin{align}
	\partial_t \phi^{(1)}_i &=  D_i \nabla^2 \phi^{(1)}_i + 2k\phi_0 - k\phi^{(1)}_i,  \qquad \text{ } \text{ } \text{ }\phi^{(1)}_A > \phi^{(1)}_I \qquad  \text{on } [0, L_\mathrm{p}] , \label{AnalyticODE1} \\
	\partial_t \phi^{(2)}_i &=  D_i \nabla^2 \phi^{(2)}_i  - k\phi^{(2)}_i ,  \qquad  \qquad \qquad  \phi^{(2)}_A < \phi^{(2)}_I \qquad \text{on }  [L_\mathrm{p}, L_\mathrm{p}+ L_\mathrm{d}].  \label{AnalyticODE2}
	\;,
\end{align}
\end{subequations}
where we omit cross-diffusion for now.
We furthermore impose matching conditions on $L_\mathrm{p}$, continuity, and periodicity
\begin{subequations}
\label{Matching_conditions} 
\begin{align}
	\phi_i^{(1)}(L_\mathrm{p})  &=  \phi_i^{(2)}(L_\mathrm{p})  ,  \\ 
	\partial_x \phi_i^{(1)}(L_\mathrm{p})  &=  \partial_x \phi_i^{(2)}(L_\mathrm{p})  . 
\end{align} 
\end{subequations}
We solve \Eqsref{AnalyticODE1} and \eqref{AnalyticODE2} in the steady state. The solutions in both domains are readily found in terms of hyperbolic functions. For simplicity we constructed them such that $\partial \phi_i^{(1)} (L_\mathrm{p}/2) = \partial \phi_i^{(2)} (L_\mathrm{p} + L_\mathrm{d}/2) = 0$,
\begin{subequations}
	\label{Sol_Ideal}
\begin{align}
	\phi_A^{(1)} &= 2 c_1 \cosh(\xi_A^{-1}(x- L_\mathrm{p}/2)) + 2\phi_0 ,   \label{Sol_Ideal1}  \\ 
	\phi_I^{(1)} &= 2 c_2 \cosh(\xi_I^{-1}(x- L_\mathrm{p}/2)) + 2\phi_0 ,   \\ 
	\phi_A^{(2)} &= 2 c_3 \cosh(\xi_A^{-1}(x- (L_\mathrm{p}+ L_\mathrm{d}/2)))  ,    \\
	\phi_I^{(2)} &= 2 c_4 \cosh(\xi_I^{-1}(x- (L_\mathrm{p} + L_\mathrm{d}/2)))  ,   \label{Sol_Ideal4}
\end{align}
\end{subequations}
where $\xi_A = \sqrt{D_A/k}$ and $\xi_I = \sqrt{D_I/k}$ are reaction-diffusion length scales.
This leaves $4$ unknown coefficients, $c_1, c_2, c_3, c_4$.
Imposing the matching conditions \eqref{Matching_conditions} gives 
\begin{subequations}
	\label{Sol_IdealSecond}
\begin{align}
	c_1 &= - \phi_0  \sinh^{-1}\left(\frac{1}{2} \xi_A^{-1} (L_\mathrm{p}+ L_\mathrm{d})\right)
		\sinh\left(\frac{\xi_A^{-1} L_\mathrm{d}}{2}\right),  \label{Sol_Ideal5}  \\ 
	c_2 &= - \phi_0  \sinh^{-1}\left(\frac{1}{2} \xi_I^{-1} (L_\mathrm{p}+ L_\mathrm{d})\right)
		\sinh\left(\frac{\xi_I^{-1} L_\mathrm{d}}{2}\right) ,  \\ 
	c_3 &=  \phi_0  \sinh^{-1}\left(\frac{1}{2} \xi_A^{-1} (L_\mathrm{p}+ L_\mathrm{d})\right)
		\sinh\left(\frac{\xi_A^{-1} L_\mathrm{d}}{2}\right) ,  \\ 
	c_4 &=   \phi_0  \sinh^{-1}\left(\frac{1}{2} \xi_I^{-1} (L_\mathrm{p}+ L_\mathrm{d})\right)
		\sinh\left(\frac{\xi_I^{-1} L_\mathrm{d}}{2}\right) . \label{Sol_Ideal8} 
\end{align}
\end{subequations}
To impose a condition on the length scales in the production and degradation regimes we use that the integrated production flux has to equal the integrated degradation flux so ensure there is no net accumulation or depletion. This ensures the system reaches a steady state. 
\begin{align}
	\int_{0}^{L_\mathrm{p} }2 k \phi_0 - k \phi_i^{(1)} - \int_{L_\mathrm{p}}^{L_\mathrm{p} + L_\mathrm{d}  }  k \phi_i^{(1)}  = 0 , \label{eqn_conservation}
\end{align}
Substituting the solution found above imposes $L_p = L_d$.  \Eqsref{Sol_Ideal} together with the conservation \eqref{eqn_conservation} admit periodic  solutions for any $L_\mathrm{p} = L_\mathrm{d}$. This reflects that multple pattern wavelengths are steady states and the system reflects metastability. Hence, the actual pattern length scale cannot be fixed analytically in this limit but must instead be determined from numerical simulations. 

\subsection{Cross diffusion}
 \label{app_Analytical_crossDiff}
 
\begin{figure*}
	\centering
	\includegraphics[width=\linewidth]{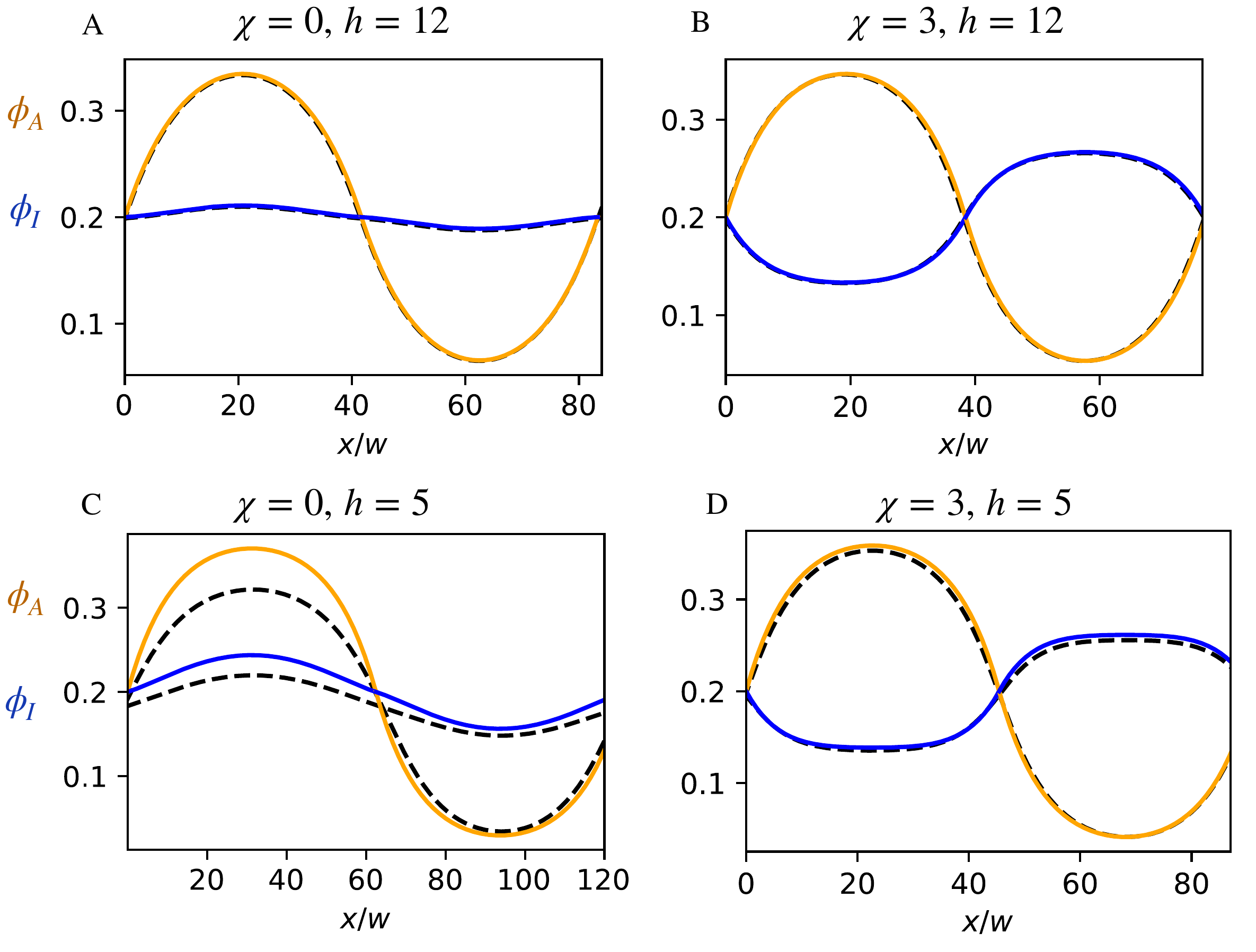}  
	\caption{Comparison of analytical results with direct numerical simulation of the reaction-diffusion equations including cross-diffusion (Eq. 5 in the main text).
	(A)~$\chi = 0$, (B)~$\chi = 3$, with $h = 12$, $D_I/D_A = 10$, and $k = 0.01$. (C)~$\chi = 0$, (D)~$\chi = 3$, with $h = 5$, $D_I/D_A = 10$, and $k = 0.01$ Analytical profiles for $\phi_A$ and $\phi_I$ are shown in orange and blue for activator and inhibitor respectively.  Numerical simulation with explicit cross diffusion is shown in black dashed.	}
		\label{fig:CompareAnalyticsH12}
\end{figure*}

We now repeat the above calculation including cross-diffusion, which couples the differential equations for the two species.
We will approach this problem by diagonalising the equations using the eigenvectors and eigenvalues of the diffusion matrix.
Still using the separation into production and degradation zones, we write
\begin{subequations}
	\label{AnalyticODE}
\begin{align}
	\partial_t  \begin{pmatrix}
	\phi_A^{(1)} \\ 
	\phi_I^{(1)}
	\end{pmatrix} &=    \begin{pmatrix}
	D_{AA} & D_{AI} \\ 
	D_{IA} & D_{II} \end{pmatrix}   \begin{pmatrix}
	\phi_A^{(1)} \\ 
	\phi_I^{(1)}
	\end{pmatrix} + 2k \begin{pmatrix}
	\phi_0 \\ 
	\phi_0
	\end{pmatrix}  - k  \begin{pmatrix}
	\phi_A^{(1)} \\ 
	\phi_I^{(1)}
	\end{pmatrix}  \qquad \text{ } \text{ } \text{ }\phi^{(1)}_A > \phi^{(1)}_I \qquad  \text{on } [0, L_\mathrm{p}] \label{AnalyticODE3} \\
	\partial_t  \begin{pmatrix}
	\phi_A^{(2)} \\ 
	\phi_I^{(2)}
	\end{pmatrix} &=    \begin{pmatrix}
	D_{AA} & D_{AI} \\ 
	D_{IA} & D_{II} \end{pmatrix}   \begin{pmatrix}
	\phi_A^{(2)} \\ 
	\phi_I^{(2)}
	\end{pmatrix}  - k  \begin{pmatrix}
	\phi_A^{(2)} \\ 
	\phi_I^{(2)}
	\end{pmatrix}  \qquad  \qquad \qquad \qquad   \phi^{(2)}_A < \phi^{(2)}_I \qquad  \text{on } [L_\mathrm{p}, L_\mathrm{p}+ L_\mathrm{d}] \label{AnalyticODE4} 
\end{align}
\end{subequations}
We determine the eigenvectors and eigenvalues of the diffusion matrix, $(\lambda_-, v_-), (\lambda_+, v_+)$.
To reflect  that the inhibitor is the fast diffusing species, we order $\lambda_+ > \lambda_-$. The eigenvalues are  
\begin{align}
\lambda_{\pm} = \frac{1}{2} (D_{AA} + D_{II}) \pm \sqrt{D_{AA}^2 + 4D_{AI} D_{IA} - 2D_{AA}D_{II} + D_{II}^2}. 
\end{align}
Using the matrix of eigenvectors $V =  \begin{pmatrix}
v_-^1& v_+^1  \\ 
v_-^2& v_+^2  \end{pmatrix} $,  
we can then diagonalise the problem using that 
\begin{align}
V^{-1} D V = {\rm Diag}(\lambda_-, \lambda_+) , 
\end{align}
where $D$ is the diffusivity matrix in \Eqsref{AnalyticODE}.
We furthermore  define rotated states
 \begin{equation}
\begin{pmatrix}
\tilde{\phi}_A^{(1)} \\ 
\tilde{\phi}_I^{(1)}
\end{pmatrix}  =  V^{-1} \begin{pmatrix}
\phi_A^{(2)} \\ 
\phi_I^{(2)}
\end{pmatrix} \qquad \begin{pmatrix}
\tilde{\phi}_A^{(2)} \\ 
\tilde{\phi}_I^{(2)}
\end{pmatrix}  = V^{-1 } \begin{pmatrix}
\phi_A^{(2)} \\ 
\phi_I^{(2)}
\end{pmatrix}
\end{equation}
and the rotated homogenous term reads
\begin{equation}
	\begin{pmatrix}
	\phi_0^A \\ 
	\phi_0^I
	\end{pmatrix} = V^{-1} \begin{pmatrix}
	\phi_0 \\ 
	\phi_0
	\end{pmatrix} 
\end{equation}
In this basis, \Eqsref{AnalyticODE} become diagonal
\begin{subequations}
\begin{align}
	\partial_t  \begin{pmatrix}
	\tilde{\phi}_A^{(1)} \\ 
	\tilde{\phi}_I^{(1)}
	\end{pmatrix} &=    \begin{pmatrix}
	\lambda_- & 0 \\ 
	0 & \lambda_+\end{pmatrix}   \begin{pmatrix}
	\tilde{\phi}_A^{(1)} \\ 
	\tilde{\phi}_I^{(1)}
	\end{pmatrix} + 2k \begin{pmatrix}
	\phi_0^A \\ 
	\phi_0^I
	\end{pmatrix} - k  \begin{pmatrix}
	\tilde{\phi}_A^{(1)} \\ 
	\tilde{\phi}_I^{(1)}
	\end{pmatrix}  \qquad \text{ } \text{ } \text{ }\phi^{(1)}_A > \phi^{(1)}_I \qquad  \text{on } [0, L_\mathrm{p}] \label{AnalyticODE5} \\
	\partial_t  \begin{pmatrix}
	\tilde{\phi}_A^{(2)} \\ 
	\tilde{\phi}_I^{(2)}
	\end{pmatrix} &=    \begin{pmatrix}
	\lambda_-& 0 \\ 
	0& \lambda_+ \end{pmatrix} 
	\begin{pmatrix}
	\tilde{\phi}_A^{(2)} \\ 
	\tilde{\phi}_I^{(2)}
	\end{pmatrix}  - k  \begin{pmatrix}
	\tilde{\phi}_A^{(2)} \\ 
	\tilde{\phi}_I^{(2)}
	\end{pmatrix}  \qquad  \qquad \qquad \qquad   \phi^{(2)}_A < \phi^{(2)}_I \qquad  \text{on } [L_\mathrm{p}, L_\mathrm{p}+ L_\mathrm{d}] \label{AnalyticODE6}
\end{align}
\end{subequations}
In the diagonal basis, the solution is now the same as for the ideal case and we can thus read of the solution from \Eqsref{Sol_Ideal} and  \Eqsref{Sol_IdealSecond} leading to
\begin{subequations}
\begin{align}
	\tilde{\phi}_A^{(1)} &= 2 c_1 \cosh(\xi_A^{-1}(x- L_\mathrm{p}/2)) + 2\phi_0^A ,   \label{AnalyticSol1} \\ 
	\tilde{\phi}_I^{(1)} &= 2 c_2 \cosh(\xi_I^{-1}(x- L_\mathrm{p}/2)) + 2\phi_0^I  , \\ 
	\tilde{\phi}_A^{(2)} &= 2 c_3 \cosh(\xi_A^{-1}(x- (L_\mathrm{p}+ L_\mathrm{d}/2))) ,    \\
	\tilde{\phi}_I^{(2)} &= 2 c_4 \cosh(\xi_I^{-1}(x- (L_\mathrm{p} + L_\mathrm{d}/2)))   , 
\end{align}
\end{subequations}
where $\xi_A = \sqrt{D_A/k}$ and $\xi_I = \sqrt{D_I/k}$ are reaction-diffusion length scales. $k$ is the reaction rate and $D_I$ and $D_A$ are the diffusion constants for $\phi_A, \phi_I$ that enter in the diffusivity matrix \eqref{AnalyticODE3}. 
The matching conditions fix $c_1, c_2, c_3, c_4$ to be 
\begin{subequations}
\begin{align}
	c_1 &= - \phi^A_0  \sinh^{-1}\left(\frac{1}{2} \xi_A^{-1} (L_\mathrm{p}+ L_\mathrm{d})\right)
		\sinh\left(\frac{\xi_A^{-1} L_\mathrm{d}}{2}\right),   \\ 
	c_2 &= - \phi^I_0  \sinh^{-1}\left(\frac{1}{2} \xi_I^{-1} (L_\mathrm{p}+ L_\mathrm{d})\right)
		\sinh\left(\frac{\xi_I^{-1} L_\mathrm{d}}{2}\right) ,  \\ 
	c_3 &=  \phi^A_0  \sinh^{-1}\left(\frac{1}{2} \xi_A^{-1} (L_\mathrm{p}+ L_\mathrm{d})\right)
		\sinh\left(\frac{\xi_A^{-1} L_\mathrm{d}}{2}\right) ,  \\ 
	c_4 &=   \phi^I_0  \sinh^{-1}\left(\frac{1}{2} \xi_I^{-1} (L_\mathrm{p}+ L_\mathrm{d})\right)
		\sinh\left(\frac{\xi_I^{-1} L_\mathrm{d}}{2}\right) .  
\end{align}
\end{subequations}
Our final result of interest is then the solution in physical space, which we obtain by rotating the solution back.
The solution then becomes a linear combination of the $\tilde{\phi}$ fields,
\begin{equation}
	\begin{pmatrix}
	\phi_A^{(1)} \\ 
	\phi_I^{(1)}
	\end{pmatrix}  =  V^{} \begin{pmatrix}
	\tilde{\phi}_A^{(2)} \\ 
	\tilde{\phi}_I^{(2)}
	\end{pmatrix}  , \qquad \begin{pmatrix}
	\phi_A^{(2)} \\ 
	\phi_I^{(2)}
	\end{pmatrix}  = V^{ } \begin{pmatrix}
	\tilde{\phi}_A^{(2)} \\ 
	\tilde{\phi}_I^{(2)}
	\end{pmatrix} \label{AnalyticSol3} . 
\end{equation}
\figref{fig:CompareAnalyticsH12} shows the analytical solutions compared to a numerical simulation  for $\chi = 0, 3$ at $h = 12$.
Even if $h$ is not very large, the limit used in the approximation ($h \to \infty$) does a reasonably good agreement (\figref{fig:CompareAnalyticsH12} shows $h = 5$). 

\begin{figure*} 
	\centering
	 \includegraphics[width=\linewidth]{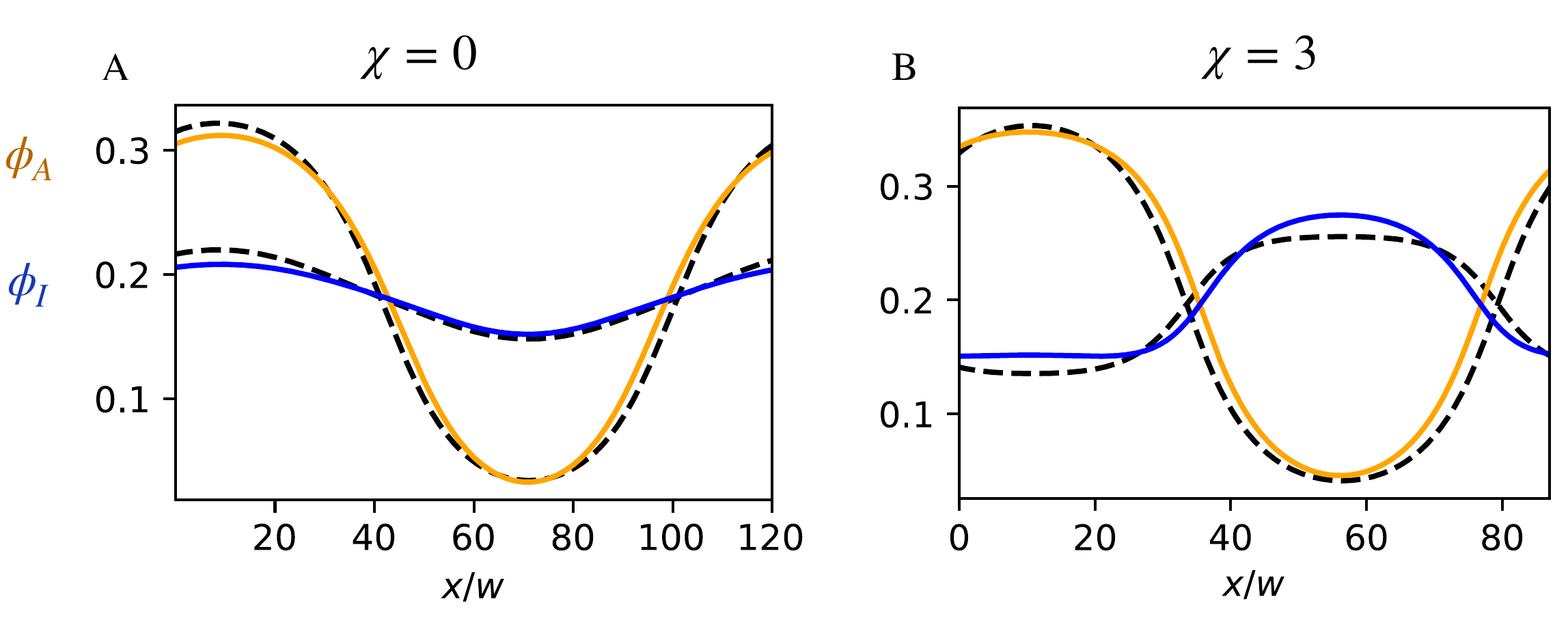}  
	\caption{Comparison of full model Eq 1 (orange/blue)  in main text  to model with explicit cross diffusion Eq 5 (black dashed). (A) $\chi = 0$, (B) $\chi = 3$. Agreement is reasonably good and the model with cross diffusion reproduces all relevant features of the patterns. Parameters are $h = 5 $, $k = 0.01$ in the limit $\Delta \mu \to \infty$.   }
		\label{fig:CompareCrossToFull}
\end{figure*}

\subsection{Role of  solvent mobility }

In the linearisation of the non-ideal diffusivity, we considered an infinitely fast solvent. The general expression for a mobility matrix for non-ideal diffusion is \cite{ZwickerPaulinTerBurg2025}
\begin{align}
\partial_t \phi_i  = \nabla \cdot (\Lambda_{ij} \nabla \mu_i), 
\end{align}
where 
\begin{align}
\Lambda_{ij} = \lambda_i \phi_i \delta_{ij} - \frac{\lambda_i \lambda_j \phi_i \phi_j}{ \sum_{k = A, I, S} \lambda_k \phi_k}, 
\end{align}
accounts for the solvent mobility $\lambda_S$.
One can then expand the fields around mean values $\bar\phi_i$ using perturbations $\delta \phi_i(x)$,
\begin{align}
	\phi_i(x) = \bar{\phi_i} +  \delta \phi_i(x) . 
\end{align}
We consider $ \bar{\phi_i} = \phi_0$ for both activator and inhibitor and define $\psi = \phi_0/(1- 2\phi_0)$.
Inserting this into $\nabla \cdot (\Lambda_{ij} \nabla \mu_i)$, using $\chi=0$ for simplicity, and expanding to first order in $\nabla \cdot (\Lambda_{ij} \nabla \mu_i)$, we find
\begin{align}
\partial_t \delta \phi_i  = D_{ij}  \nabla^2 \cdot ( \delta \phi_i ) 
\end{align}
where $D_{ij}$ is  
\begin{equation}
D_{ij} = \frac{1}{\lambda_S + (\lambda_A + \lambda_I) \phi_0 - 2 \lambda_S \phi_0}
\begin{pmatrix} 
	\lambda_A (\lambda_S + \lambda_I \phi_0 - \lambda_S \phi_0) &
	\lambda_A (-\lambda_I + \lambda_S) \phi_0
	\\[12pt]
	\lambda_I (-\lambda_A + \lambda_S) \phi_0 &
	\lambda_I (\lambda_S + \lambda_A \phi_0 - \lambda_S \phi_0)
\end{pmatrix} . 
\end{equation}
In the limit of an infinitely fast solvent, $\lambda_S \to \infty$, this reduces to 
\begin{equation}
\begin{pmatrix} 
(1+ \psi) \lambda_A& \psi \lambda_A  \\
\psi \lambda_I& (1+ \psi) \lambda_I
\end{pmatrix}, 
\end{equation}
which is consistent with Eq.~(6) in the main text for $\chi=0$.

To understand how the solvent mobility affects cross diffusion, we investigate the simple case $\lambda_A=\lambda_I=\lambda$, where we find
\begin{equation}
D_{ij} = \frac{1}{\lambda_S + 2\lambda \phi_0 - 2 \lambda_S \phi_0}
\begin{pmatrix} 
	\lambda (\lambda_S + \lambda \phi_0 - \lambda_S \phi_0) &
	\lambda (-\lambda + \lambda_S) \phi_0
	\\[12pt]
	\lambda (-\lambda + \lambda_S) \phi_0 &
	\lambda (\lambda_S + \lambda \phi_0 - \lambda_S \phi_0)
\end{pmatrix} , 
\end{equation}
which shows that the diagonal terms dominate the off-diagonal terms for small $\lambda_S$.
This suggest that influence of cross-diffusivity decreases for less mobile solvents.

\section{Energy input equals dissipation}
\label{app_energyVsDiss}

\begin{figure*}
	\centering
	\includegraphics[width=\linewidth]{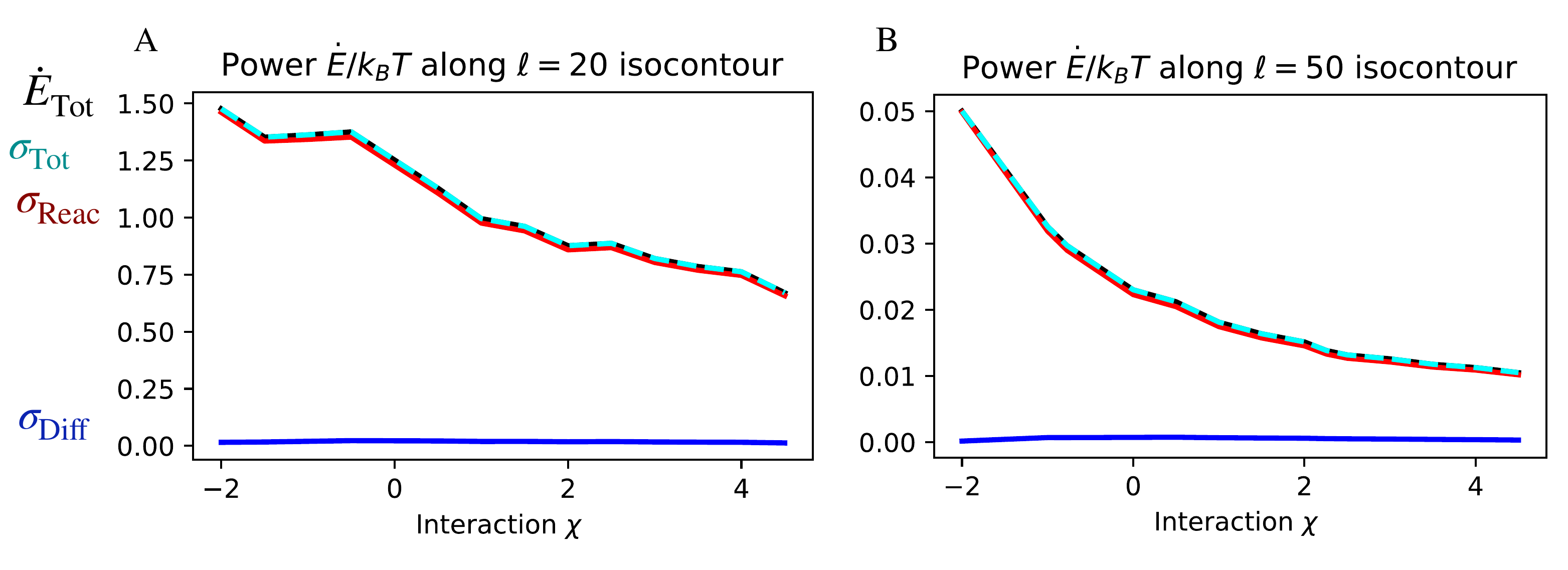}  
	\caption{Total power $\dot{E}$ determined from Eq. (4) in the main text (black) compared to total dissipation measured via \Eqref{power_eqn} (cyan-dashed), for $\Delta \mu = 5$ (panel A) and $\Delta \mu = 1$ (panel B).
	We also show the dissipation via reactions (red) and diffusion (blue) measured via \Eqref{power_eqn_reac_diff}.
	}
		\label{fig_app_energyDiff}
\end{figure*}

In the steady state. the total power $\dot{E}$ as given in Eq. 4 of the main text is equal to the total dissipation $\sigma^\mathrm{tot}$, which consists of a contribution of the diffusion and reactions,
\begin{align}
	\sigma^{{\rm tot}}   &= \frac{1}{V}  \sum_{i = A, I} \int  \bigl[\sigma ^{\rm diff}_i + \sigma  ^{\rm reac}_i  \bigl]  \mathrm{d}V , \label{power_eqn}
\end{align}
where 
\begin{subequations}
\begin{align}
	 \sigma^{\rm diff}_i    &= D_i \phi_i (\nabla \mu_i)^2  ,  \\
	\sigma ^{\rm reac}_i  &=  \sP_i( - \bar{\mu}_A + \Delta \mu)  + \sD_i(-\bar{\mu}_A - \Delta \mu)  .\label{power_eqn_reac_diff}
\end{align}
\end{subequations}
 \figref{fig_app_energyDiff} compares the energy (black), to the total dissipation as given in \eqref{power_eqn} (cyan dashed), showing excellent agreement.
 Furthermore, \Eqref{power_eqn} allows us to measure dissipation due to reactions (red) and dissipation due to diffusion (blue) separately, showing that most energy is dissipated by reactions.

\section{Homogeneous state as a function of $\Delta \mu$}

\begin{figure}
	\includegraphics[width=\columnwidth]{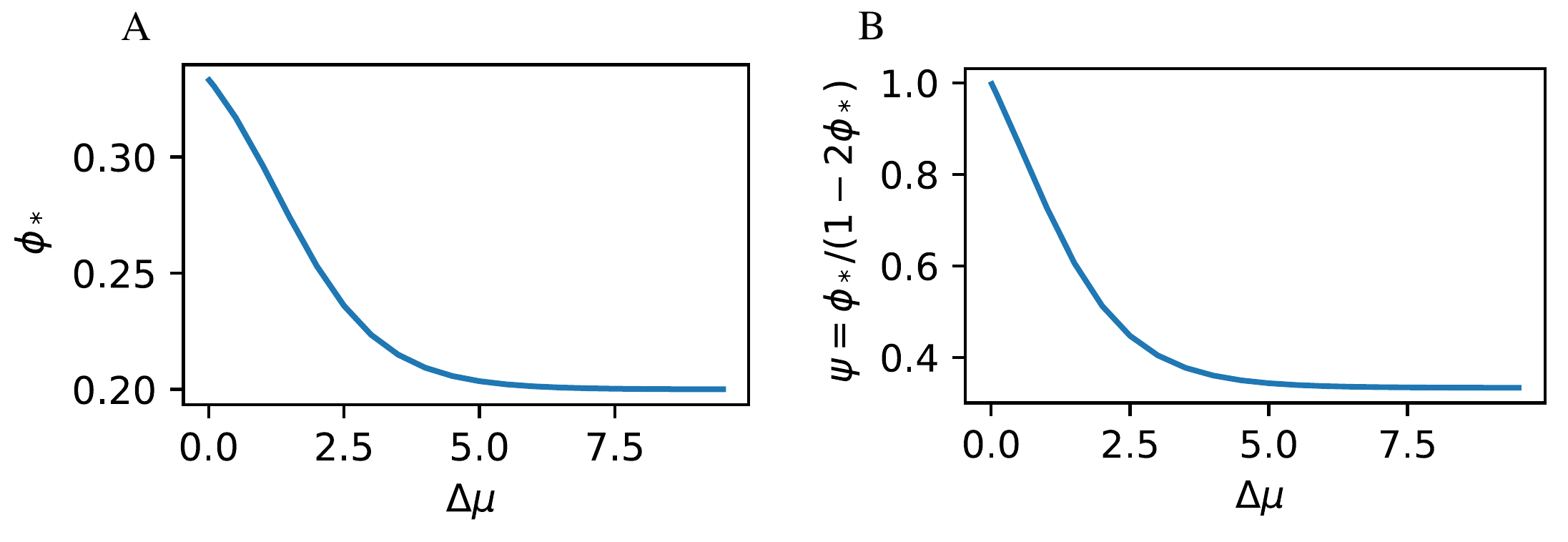}
	\caption{  (A) The fraction $\phi_*$ in the homogeneous stationary state as a function of driving strength $\Delta \mu$.
	(B) Corresponding function $\psi = \phi_*/(1 - 2\phi_*)$ as a function of $\Delta \mu$.
	Additional model parameters are $\chi = 0$ and $\phi_0 = 0.2$. 
	}
	\label{fig:HomState}
\end{figure}

\figref{fig:HomState}A shows the fraction $\phi_*$ of the homogeneous state at chemical equilibrium, imposed by the condition $s(\phi_*) = 0$.
This condition gives an implicit equation that depends on $\Delta \mu, \chi, \phi_0$ that can be solved numerically. 
\figref{fig:HomState}B shows the corresponding parameter $\psi = \phi_*/(1 - 2\phi_*)$ that enters in the diffusion matrix given by Eq. 6 in the main text.
Both quantities decrease with stronger activity~$\Delta\mu$. 

\bibliographystyle{apsrev4-1}
\bibliography{references}